\documentclass[aps,prl,twocolumn,showpacs]{revtex4}
\usepackage{epsfig}
\usepackage{graphicx}
\usepackage{amsfonts}
\usepackage[figuresright]{rotating}
\usepackage{amssymb}
\usepackage{amsmath}
\usepackage{dcolumn}
\usepackage{bm}

\def\be{\begin{equation}} \def\ee{\end{equation}}
\def\bea{\begin{eqnarray}} \def\eea{\end{eqnarray}}

\begin{document}
\title{Exotic many-body physics with large-spin Fermi gases}
\author{Congjun Wu}
\affiliation{Department of Physics, University of California,
San Diego, CA92093}
\begin{abstract}
The experimental realization of quantum degenerate cold Fermi gases 
with large hyperfine spins opens up a new opportunity for exotic
many-body physics.
\end{abstract}
\maketitle 

A Viewpoint on: 
{\it Realization of an $SU (2) \times SU (6)$ System of Fermions 
in a Cold Atomic Gas} by Shintaro Taie, Yosuke Takasu, Seiji Sugawa, Rekishu Yamazaki, Takuya Tsujimoto, Ryo Murakami and Yoshiro Takahashi,
Phys. Rev. Lett. {\bf 105}, 190401 (2010).

Cold atom physics has become a new frontier, both for
the study of known condensed matter phenomena such
as superfluid-Mott insulator transitions, and as a way of
creating novel quantum phases. Many atoms have large
spins and can be stored and investigated in optical traps
and lattices. Thus recent experimental progress on the
large-spin ultracold Fermi gases provides an exciting
opportunity to investigate exotic many-body physics, as
reported in two recent articles published in {\it Physical Review
Letters}. In one case, Desalvo et al. \cite{desalvo2010} have cooled
down the system of $^{87}$Sr ($F = I = 9/2$, where $I$ is the
nuclear spin and $F$ is the total hyperfine spin) to quantum
degeneracy (that is, a situation where the atoms occupy
the lowest possible energy states limited by Pauli
exclusion). Now, Shintaro Taie and colleagues at Kyoto
University, Japan, and the Japan Science and Technology
Agency report their success in cooling of $^{173}$Yb ($F =
I = 5/2$) \cite{taie2010}. Both $^{87}$Sr and $^{173}$Yb
have an alkaline earth-like atomic structure with all electron shells filled;
thus their hyperfine spins completely come from nuclear
spins. Since nuclei are deep within the atom, interactions
among atoms are nearly spin-independent. As
a result, we now have the opportunity to explore, in an
atomic system, complex many-body physics of the sort
usually only seen in high-energy physics.

\begin{figure}[htb]
\includegraphics[width=\linewidth]{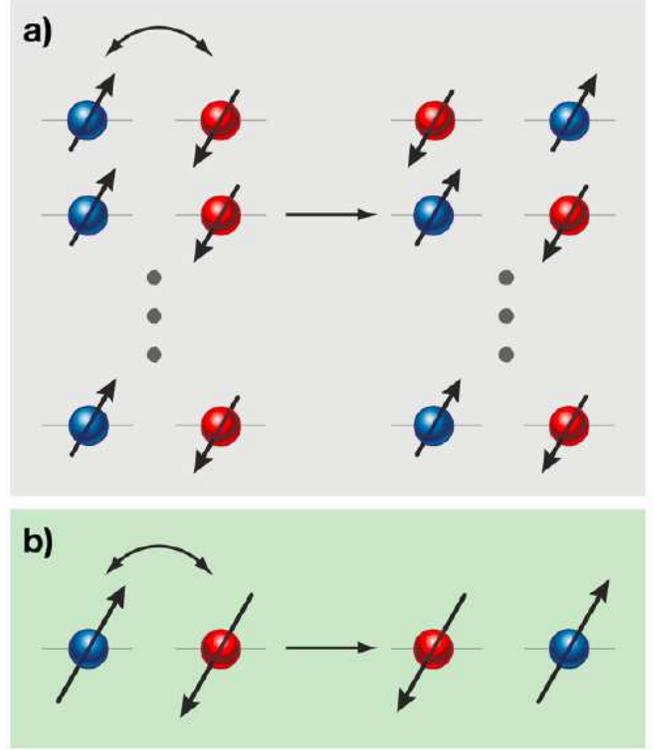}
\caption{(a) In solid-state systems with large total spin on each
site, spin fluctuations occur by super exchange, where, at the
leading order, one spin-up is exchanged for a spin-down electron.
The size of the fluctuations of $S_z$ can only be $\pm1$ in a total
spin of $S$, so the fluctuations are relatively small at large S.
(b) In cold-atom Fermi systems with large hyperfine spin, the
atom moves as an entire object. In this case, the fluctuations of
$S_z$ can be a much larger value of $\pm S$.
}
\label{fig:spin}
\end{figure}

For these atomic systems, all the $2F + 1$ hyperfine
levels are equivalent in terms of their participation in
atom-atom interactions, which gives rise to the so-called
$SU(N)$ symmetry, with $N = 2F + 1$, in the language of
symmetry groups. Mathematicians and physicists know
$SU(N)$ as the special unitary group of degree $N$, which
can be represented by a set of $N \times N$ unitary matrices
with determinant 1. The purpose of such groups
is to capture the symmetry properties that govern the
behavior of various particles. The matrices transform
a vector (representing a particle or state) in an abstract
$N$-dimensional space into another vector (or particle).
$SU(2)$, for example, appears in the physics of spin-$1/2$
particles like electrons, which can be spin up or spin
down. The $SU(3)$ group is found in quantum chromodynamics—
the physical theory of quarks and gluons—
in which there are three particle “colors.” Higher
dimensional groups, such as $SU(N)$, where $N = 2F +1$,
could then represent particles of hyperfine spin $F$. For
instance, $^{173}$Yb with spin $= 5/2$ has six components with
$F_z$ equal to  $-5/2,-3/2,-1/2,+1/2,+3/2$, and $+5/2$,
requiring an $SU(6)$ symmetry group. Taie {\it et al.} have
gone further, however, loading the $^{173}$Yb atoms into
three-dimensional optical lattices together with atoms
of its spin-$1/2$ isotope, $^{173}$Yb. This mixture of spin
species is then a realization of an $SU(6)\times SU(2)$ system
\cite{taie2010}, where the direct product symbol $\times$ means that the
symmetry transformation in one isotope sector is independent
of that in the other.

In general, cold Fermi systems with large hyperfine
spins have aroused a great deal of theoretical interest.
Researchers have studied the rich structure of
collective excitations in Fermi liquid theory \cite{yip1999} and
the Cooper pairing patterns of large spin fermions \cite{ho1999}. 
Considerable progress has been made in the simplest
large-hyperfine-spin systems with $F = 3/2$ (e.g.,
$^{132}$Cs, $^9$Be,$^{135}$Ba, $^{137}$Ba, $^{201}$Hg)
\cite{wu2003,wu2006}, which also include
non-alkaline-earth atoms with nonzero electron spins
due to partially filled electron shells. In this case, the
$SU(N)$ symmetry is not generic. Instead, a hidden
$SO(N)$ symmetry with $N = 5$ is proved without fine
tuning, where $SO(N)$ is the special orthogonal group
describing rotations in an $N$-dimensional real space.
Important effects from such as symmetry on quantum
magnetism and Cooper pairing have been studied 
\cite{wu2005,controzzi2006,wu2010}. 
Recently, $SU(N)$ models have been proposed for the
alkaline-earth fermion atoms, whose effects on quantum
magnetism have been explored based on the field-
theoretical method of large-$N$ \cite{gorshkov2010,hermele2009}. 
The possible ferromagnetic
states have also been studied for the $SU(6)$
symmetric system of $^{173}$Yb\cite{cazalilla2009}.

Why are such large-spin Fermionic cold-atom systems
so important? The reason is the fundamental difference
between large-hyperfine-spin Fermi systems and largeelectron-
spin solid-state systems. For example, in many
transition-metal oxides, several electrons on the same
cation site combine into a composite object with a large
 electron spin $S > 1/2$ through Hund's rule coupling. (In
chemistry, Hund's rule refers to the situation when electron
shells are partially filled and the total spin of unpaired
electrons adds together.) Usually, however, large
spins are not particularly interesting compared to spin
1/2, because quantum spin fluctuations become weaker
for large spins in electron systems. This point is intuitively
illustrated in Fig. 1 (a). The intersite coupling between
large electron spins is dominated by the exchange
of a single pair of electrons. The change of $S_z$ can only
be $\pm1$, regardless of the value of $S$, thus spin fluctuations
are on the order of $1/S$. As a result, the larger the
electron spins are, the more classical is the physics they
exhibit.

In contrast, such a restriction does not exist in cold
Fermi systems with large hyperfine spins. Each atom
moves as a whole object carrying a large hyperfine spin,
as depicted in Fig. 1 (b). Exchanging cold fermions
can completely flip the total hyperfine-spin configuration.
Consequentially, as pointed out in Ref. \cite{wu2006}, quantum
fluctuations are enhanced by the large number of
hyperfine-spin components $N$, which are even stronger
compared to those in spin-$1/2$ systems. In other words,
large-hyperfine-spin cold-atom systems can exist in the
large-$N$ limit. This feature gives rise to the possibilities
of realizing exotic quantum magnetic states 
\cite{wu2005,gorshkov2010,hermele2009},
including various spin liquids and valence-bond solid
states that are difficult to stabilize in electron systems.

Large-hyperfine-spin cold fermions also bring nontrivial
features to the Cooper pairing found in superfluidity.
In conventional spin-$1/2$ systems, the only possibility
for the $s$-wave pairing is that of spin singlet, that
is, a spin-up paired with a spin-down fermion, with the
total spin zero. In the large-hyperfine-spin case, the $s$-wave
Cooper pair can take even values of the total hyperfine
spin, ranging from 0, 2 to $2S-1$ \cite{ho1999}. This results
in interesting spin dynamics of pairing condensates, and
exotic topological defects with combined vortices in the
phase and spin configurations of the pairing condensate
\cite{ho1999,wu2010}.

Another exotic physical phenomenon in large hyperfine-
spin cold fermions is the ``baryon-like''
multiple-fermion clustering instability. In an $N$-component
Fermi system with attractive interactions,
Pauliâs exclusion principle allows, at most, $N$ fermions
to form an $SU(N)$ singlet state \cite{wu2005,lecheminant2008}. 
Compared to
Cooper pairing, such a state has a lower interaction energy
but costs more kinetic energy, thus is favored when
interactions are strong. In quantum chromodynamics
(QCD), baryons, which are color singlets, are composed
of three quarks because two quarks cannot form a color
singlet for the $SU(3)$ color gauge group. It is interesting
that in spite of the huge difference of energy scales, the
large-spin cold fermions can also exhibit similar physics
to that in QCD \cite{rapp2007}.

The exploration of many-body physics with large spin
ultracold fermions is a promising research direction
of cold-atom physics. A major experimental challenge is
the precise control of interactions among atoms, which
is important to realize exotic Cooper pairing with rich
internal spin structure. To experimentally study exotic
quantum magnetism with large-spin fermions, further
cooling the temperatures below the magnetic exchange
energy scale is necessary. It is natural to expect that this
research will greatly expand our knowledge of many body
physics different from that in electron systems.

\end{document}